\documentstyle[prd,aps]{revtex}

\begin{document}

\draft

\title{BLACK HOLE SOLUTIONS WITH DILATONIC HAIR IN HIGHER CURVATURE GRAVITY}

\author{S.O.Alexeyev, M.V.Pomazanov}

\address{Department of Theoretical Physics, Physics Faculty,\\
Moscow State University, \\
Moscow 119899, RUSSIA \\
e-mail: alexeyev@grg2.phys.msu.su}

\date{\today}

\maketitle

\begin{abstract}
A new numerical integration method for examining a black hole structure
was realized. Black hole solutions with dilatonic hair of 4D low energy
effective SuperString Theory action with Gauss-Bonnet quadratic curvature
contribution were studied, using this method, inside and outside
the event horizon. Thermodynamical properties of this solution were
also studied.
\end{abstract}

\pacs{0260Lj, 0270Rw, 0425Dm, 0470Bw, 0470Dy}

\section{Introduction}

During last years a great interest for an investigation of low energy
effective SuperString Theory action in four dimensions arised.
Some researches \cite{witt,mignemi,kanti,donets}
found that the well-known solutions (such as Schwarzshild one and ets.)
were modified by the higher order curvature corrections.
They also showed that the ``no-hair'' theorem can not been applied
to those modified configurations.

The problem is to find black-hole solutions of 4D low energy String
effective action with the second order curvature corrections.
For simplicity, as a rule, researchers \cite{mignemi,kanti}
consider the bosonic part of the gravitational action
consisting of dilaton, graviton and Gauss-Bonnet (GB)
terms taken in the following form:
\begin{eqnarray}\label{eq1}
S & = & \frac{1}{16\pi} \int d^4 x \sqrt{-g}
 \biggl[ m^2_{pl} (-R+2\partial_{\mu} \phi
\partial^\mu \phi ) + \lambda \mbox{e}^{-2\phi} S_{GB} \biggr],
\end{eqnarray}
where $R$ is a scalar curvature; $\phi$ is a dilaton field;
$m_{pl}$ is the Plank mass;
$\lambda$ is the string coupling parameter.
The later describes GB contribution
($S_{GB}=R_{ijkl}R^{ijkl} - 4 R_{ij}R^{ij} + R^2 $)
to the action (\ref{eq1}).
Such configurations were partly studied \cite{mignemi}
by the perturbative analysis $O(\lambda)$ outside the event horizon
when event horizon radius $r_h \gg m_{pl}$. Authors showed the black hole
solution to be real and it provides the non-trivial dilatonic hair.
P.Kanti et all \cite{kanti} obtained the similar solution
using the non-perturbative numerical method outside the horizon.
However, a question on a solution behavior inside the event horizon
is still open.
It is generally believed \cite{frolov} that in the regions where the
spacetime curvature is just small a \underline{classical}
solution gives the main contribution to the global structure
of the spacetime. Quantum corrections may drastically modify
the spacetime properties in the case of large enough curvature.
The purpose of this paper is to study the complete black-hole solution.

Structure of our paper is the following: in Section 2
an analytical investigation of the action (\ref{eq1}) is described;
Section 3 deals with the numerical results;
Section 4 is devoted to description
of thermodynamical properties of the solution;
Section 5 contains a discussion and conclusions.
In Appendix one can find a brief description of our integration method.

\section{Analytical Investigation of the Gauss-Bonnet Action}

The aim is to find static, asymptotically flat, spherically symmetric
black-hole-like solutions. In this case the most convenient choice
of metric is the following:
\begin{equation}
ds^2 = \Delta dt^2 - \frac{\sigma^2}{\Delta} dr^2 - f^2
(d \theta^2 + \sin^2 \theta d \varphi^2),
\end{equation}
where functions $\Delta$, $\sigma$ and $f$ depend only on radial coordinate $r$.
Therefore, the scalar curvature $R$ and the GB term $S_{GB}$ have the following forms:
\begin{eqnarray}
R \sqrt{-g} \frac{1}{\sin\theta}
& = & \biggl[\frac{\Delta'f^2}{\sigma} + 4 \frac{f' \Delta f}{\sigma}\biggr]'
 -  \frac{2}{\sigma} \biggl[\Delta' f' f +\Delta (f')^2
+ \sigma^2\biggr], \\
S_{GB} \sqrt{-g} \frac{1}{\sin\theta} & = &
4\biggl[ \frac{\Delta \Delta' (f')^2 }{\sigma^3}
- \frac{\Delta'}{\sigma}\biggr]'.
\end{eqnarray}
Integrating i) by parts the GB term and $R$ term and ii) over the angle variables
in (\ref{eq1}), one can rewrite the action in somewhat more convenient
form (for the present analysis the boundary
term is not relevant and is ignored):
\begin{equation}
S=\frac{1}{2} \int dt dr \biggl[
m^2_{pl} \frac{1}{\sigma} (\Delta' f' f + \Delta (f')^2 + \sigma^2
-\Delta f^2 (\phi')^2) + 4 \mbox{e}^{-2\phi} \lambda \phi'
( \frac{\Delta \Delta' (f')^2 }{\sigma^3}
 - \frac{\Delta'}{\sigma}) \biggr].
\end{equation}
Corresponding field equations in the curvature gauge ($f(r)=r$)
are the following:
\begin{eqnarray}\label{eq2.1}
&& m^2_{pl} \sigma^2 [-r\sigma' +\sigma r^2 (\phi')^2]
 + 4 \mbox{e}^{-2\phi} \lambda \sigma (\Delta - \sigma^2) [\phi'' - 2 (\phi')^2] +
4 \mbox{e}^{-2\phi} \phi' \lambda \sigma' (\sigma^2 - 3\Delta) = 0,
\end{eqnarray}
\begin{eqnarray}\label{eq2.2}
&& m^2_{pl} \sigma^2 [ \sigma^2 + \Delta r^2 (\phi')^2 - \Delta' r - \Delta ] +
4 \mbox{e}^{-2\phi} \phi' \lambda \Delta' (\sigma^2 - 3\Delta) = 0 ,
\end{eqnarray}
\begin{eqnarray}\label{eq2.3}
&& m^2_{pl} \sigma^2 [\Delta'' r \sigma - \Delta' r \sigma' +2 \Delta' \sigma
- 2 \Delta \sigma' +2 \Delta r \sigma (\phi')^2 ] \nonumber \\
&& + 4 \mbox{e}^{-2\phi} \lambda 2 \sigma \Delta \Delta' [\phi'' - 2 (\phi')^2]
+ 4 \mbox{e}^{-2\phi} \phi' \lambda 2 [(\Delta')^2 \sigma +\Delta \Delta'' \sigma -
3\Delta \Delta' \sigma' ] = 0 ,
\end{eqnarray}
\begin{eqnarray}\label{eq2.4}
&& -2 m^2_{pl} \sigma^2 [ \Delta' r^2 \sigma \phi' + 2 \Delta r \phi' \sigma -
\Delta r^2 \phi' \sigma' + \Delta r^2 \phi'' \sigma ] \nonumber \\
&& + 4 \mbox{e}^{-2\phi} \lambda [(\Delta')^2 \sigma +\Delta \Delta'' \sigma -
3 \Delta \Delta' \sigma' - \Delta'' \sigma^3 + \Delta' \sigma' \sigma^2 ] = 0.
\end{eqnarray}
The equation (\ref{eq2.2}) can be solved analytically and it permits
to get the following expression for $\sigma$:
\begin{eqnarray}\label{eq2.5}
\sigma^2 & = & \frac{-b+\sqrt{b^2-4ac}}{2a},  \\
\mbox{where}\nonumber \\
a & = & m^2_{pl}, \qquad
b   =  m^2_{pl} (\Delta r^2 (\phi')^2 -\Delta' r -\Delta )
+ 4 \mbox{e}^{-2\phi}\lambda \phi' \Delta',  \qquad
c   =  -4 \mbox{e}^{-2\phi}\lambda \phi' (3 \Delta \Delta'). \nonumber
\end{eqnarray}
In a formal mathematical way it is necessary to put the formula (\ref{eq2.5})
to equations (\ref{eq2.1}), (\ref{eq2.3}), (\ref{eq2.4}) and to work with two unknown
functions $\Delta$ and $\phi$, but in the given situation it is more convenient
to work with the equations (\ref{eq2.1}), (\ref{eq2.3}), (\ref{eq2.4})
and three unknown functions $\Delta$, $\sigma$ and $\phi$,
verifying (\ref{eq2.2}) at every numerical integration step and writing
asymptotic expressions for all four equations.
We choose the later strategy.

Let us suppose the solutions behaviors near the event horizon to
have the following forms:
\begin{eqnarray}
\Delta & = & d_1 x + d_2 x^2 + O(x^2), \label{eq2.6} \\
\sigma & = & s_0 +s_1 x + O(x), \label{eq2.61} \\
\mbox{e}^{-2\phi} & = & \phi_0
(1 - 2 * \phi_1 x + 2 (\phi^2_1 - \phi_2) x^2 ) + O(x^2), \label{eq2.62}
\end{eqnarray}
where $x=r-r_h, \ll 1$.
Putting formula (\ref{eq2.6})-(\ref{eq2.62})
in the equations (\ref{eq2.1})-(\ref{eq2.4}),
we obtain the following equation and relations
between the expansion coefficients ($s_0$, $s_1$ and $\phi_0$ are free
parameters):
\begin{equation}\label{eq2.7}
z_1 d_1^2 + z_2 d_1 + z_3 = 0,
\end{equation}
where:
\begin{eqnarray}
z_1  = 24 \lambda^2 \phi_0^2 r_h, \qquad
z_2  = 8 \lambda^2 \phi_0^2 s_0^2 - m^4_{pl} r^4_h s_0^2,  \qquad
z_3  = m^4_{pl} r^3_h s_0^4; \nonumber
\end{eqnarray}
and other parameters are:
\begin{eqnarray}\label{eq2.71}
d_2 & = & ( - \frac{1}{2 r_h s^2_0}) [r_h d_1^2 + d_1 s_0^2 - r_h s_0 s_1 d_1 ], \\
\phi_1 & = & \frac{m_{pl}^2}{4 \lambda d_1 \phi_0} [r_h d_1 - s_0^2 ], \\
\phi_2 & = & \frac{1}{8 \lambda s_0 \phi_0}
[8 \lambda \phi^2_1 \phi_0 + 4 \lambda s_1 \phi_1 \phi_0 +
r_h^2 s_0 \phi_1^2 m_{pl}^2 - r_h s_1 m_{pl}^2 ].
\end{eqnarray}
The solution of quadratic equation (\ref{eq2.7}) exists if a value of the free
parameter $\phi_0$ satisfies the following condition:
\begin{eqnarray}
\phi_0^2 \geq \frac{7 (1+\sqrt{195}) m^4_{pl} r^4_h}{2 \lambda^2}.
\end{eqnarray}
This is a natural restriction to the solution existence of the equation
system (\ref{eq2.1})-(\ref{eq2.4}).

We are interesting in asymptotically flat solutions. Therefore,
its infinity behavior is to be as follows:
\begin{eqnarray}\label{eq2.8}
\Delta = 1 - \frac{2M}{r} + O\biggl(\frac{1}{r}\biggr); \quad
\sigma = 1 + O\biggl(\frac{1}{r}\biggr); \quad
\phi = \phi_\infty + \frac{D}{r} + O\biggl(\frac{1}{r}\biggr),
\end{eqnarray}
where ADM mass $M$ and dilaton charge $D$ can be calculated
>from the equations (\ref{eq2.3}), (\ref{eq2.4})
by the following formulae:
\begin{eqnarray}\label{eq2.9}
 \biggl( \frac{\Delta' r^2 }{\sigma} \biggr)  =  2 M & = & \int\limits_{r_h}^{\infty}
d r \frac{1}{\sigma^4 m^2_{pl} } \biggl( 2 m^2_{pl} \sigma^2 \biggl[
\Delta r \sigma' - \Delta r^2 (\phi')^2 \sigma \biggr] \nonumber \\
&& - 4 \mbox{e}^{-2\phi} \lambda 2 \sigma \Delta \Delta' r [\phi'' - 2 (\phi')^2]
- 4 \mbox{e}^{-2\phi} \phi' \lambda r [(\Delta')^2 \sigma +\Delta \Delta'' \sigma -
3\Delta \Delta' \sigma' ]  \biggr), \\
\nonumber \\
 \biggl( \frac{2 \Delta r^2 \phi'}{\sigma} \biggr)  =  D & = & \int\limits_{r_h}^{\infty}
d r \frac{1}{m^2_{pl}\sigma^4} \biggl(
 4 \mbox{e}^{-2\phi} \lambda [(\Delta')^2 \sigma +\Delta \Delta'' \sigma
 - 3 \Delta \Delta' \sigma' - \Delta'' \sigma^3 + \Delta' \sigma' \sigma^2 ] \biggr).
\end{eqnarray}
The dependencies of ADM mass $M$ and dilaton charge $D$
versus the event horizon radius $r_h$ are shown in Tab.1.

\section{Numerical Results}

For integrating inside the event horizon a method based on integrating
over an additional parameter was used. It is briefly described
in Appendix. Here we present the main results.

The integration was done from the event horizon to the infinity,
then, using data obtained, from the infinity to the horizon and, further,
inside the horizon.  The results of our calculation are shown in Fig.1.
It presents the dependence of metric functions $\Delta$ and $\sigma$ and the
dilaton function $\phi$ versus $r$. Parameter $r_h$ was allowed
to change within wide range from $3.5$ to $100.0$ Plank unit values.
Fig.2 represents 3D plot of the dependence of the metric function $\Delta(r)$
and dilaton function $\exp(-2\phi(r))$
versus $r_h$. As one can see from Fig.1 and Fig.2 the behaviors of
$\Delta$, $\sigma$ and $\phi$ outside the horizon have the usual forms,
as obtained in \cite{mignemi,kanti}, and look like
the standard Schwarzshild solution (ADM mass values $M$ ($r_h \simeq 2M$)
illustrate this, see Table 1).
Under the horizon $r < r_h$ solution exist only till
the value $r=r_s$.
Another solution branch begins from the value $r_s$, but it
exists only till the ``singular'' horizon $r_x$.
The asymptotic behavior of both solution branches near the position $r_s$
can be described by the following formulae, using the only smooth
function $\sigma$ as an independent variable:
\begin{eqnarray}\label{ew}
\Delta & = & d_s + d_2 x^2 + O(x^2) ; \nonumber \\
r & = & r_s + r_2 x^2 + O(x^2); \\
\exp(-2\phi) & = & \phi_s (1- 2 f_2 x^2 ) + O(x^2), \nonumber
\end{eqnarray}
where $x=\sigma - \sigma_s \ll 1$.
Free parameters are the following: $\sigma_s$, $\phi_s$, $r_s$.
Other expansion coefficients ($d_s$, $f_2$, $d_2/r_2$)
can be calculated from the following three equations:
\begin{eqnarray}\label{eq2.10a}
&& d_2 = f_2,
\end{eqnarray}
\begin{eqnarray}\label{eq2.10}
&& m^2_{pl} \sigma^2_s \biggl[ \sigma^2_s + d_s r^2_s \biggl( \frac{d_2}{r_2} \biggr)^2
- r_s \frac{d_2}{r_2} - d_s \biggr] +4 \phi_s \lambda \biggl( \frac{d_2}{r_2} \biggr)^2
(\sigma^2_s -3 d_s) = 0,
\end{eqnarray}
\begin{eqnarray}\label{eq2.11}
&& m^6_{pl} \sigma^6_s d_s r^4_s +
4 \phi_s \lambda m^4_{pl} 4 \frac{d_2}{r_2} \sigma^4_s d_s^2 r^3_s \nonumber \\
&& + (4 \phi_s \lambda)^2 m^2_{pl} \sigma^2_s \biggl[ 4 \biggl( \frac{d_2}{r_2} \biggr)^2
d^3_s r^2_s + (d_s + r_s \frac{d_2}{r_2})(d_s - \sigma^2_s )^2 \biggr]
 + (4 \phi_s \lambda)^3 3 d_s \biggl( \frac{d_2}{r_2} \biggr)^2
(d_s - \sigma^2_s)^2 = 0.
\end{eqnarray}

The equation (\ref{eq2.10}) represents the asymptotic form of the equation
(\ref{eq2.2}).
The formula (\ref{eq2.10a}) represents the asymptotic form of the equation
(\ref{eq2.3}) and the rest of equations (\ref{eq2.1}), (\ref{eq2.4})
reduce into an identity.
The equation (\ref{eq2.11}) is the consequence of the system
(\ref{a.1}) (see Appendix) because according to the existence theorem (see Appendix)
the system (\ref{a.1})
has a single solution only in the
case of its main discriminant to be not equal to zero.
In the case the zero main discriminant in some point of the solution
trajectory, the uniqueness of the solution (\ref{a.1})
will be violated.
So, equation (\ref{eq2.11})
is just that condition in the $r_s$, where \underline{two}
solutions exist. ``Curvature invariant'' $R_{ijkl}R^{ijkl}$ in
our metric parametrization is equal to:
\begin{eqnarray}\label{eq2.12}
R_{ijkl}R^{ijkl} & = &
4 \frac{\Delta^2}{\sigma^4 r^4} +
8 \frac{\Delta^2 (\sigma')^2}{\sigma^6 r^2} -
8 \frac{\Delta}{\sigma^2 r^4} -
8 \frac{\Delta \Delta' \sigma'}{\sigma^5 r^2}
+ \frac{(\Delta'')^2}{\sigma^4} +
4 \frac{(\Delta')^2}{\sigma^4 r^2} -
2 \frac{\Delta'' \Delta' \sigma'}{\sigma^5} +
\frac{(\Delta')^2 (\sigma')^2}{\sigma^6} +
\frac{4}{r^4} \nonumber \\
& = & \frac{1}{x^2} \ \biggl( \frac{4 d_s}{r_2 r^2_s \sigma^6_s} +
\frac{d_2}{2 r^2_2 \sigma^6_s} \biggr) + O\biggl(\frac{1}{x} \biggr)
\rightarrow \infty .
\end{eqnarray}
According to the classification from \cite{ellis}, $r_s$ represents
pure scalar singularity. It is necessary to note,
as we tested, that the weak
energy condition, dominant one and strong one are realized at
 the value $r=r_s$
\cite{hawking}.

The most interesting result is the existence of the second solution branch
within the ``singular'' horizon $r_x$. The position of $r_x$ is situated inside
the main horizon $r_h$. The asymptotic behavior of $\Delta$, $\sigma$
and $\phi$ near $r_x$ are the following:
\begin{eqnarray}\label{eq2.13}
\Delta  =  d_1 x +d_2 x^{\frac{3}{2}} + O (x^{\frac{3}{2}}),  \qquad
\sigma  =  \sigma_0 +\sigma_1 x^{\frac{1}{2}} + O (x^{\frac{1}{2}}),  \qquad
\phi  =  \phi_{x} + \phi_1 x + \phi_2 x^{\frac{3}{2}} + O (x^{\frac{3}{2}}), \qquad
\nonumber
\end{eqnarray}
where $x = r_x - r \ll 1$, and free parameters are $\sigma_0$, $\sigma_1$,
$\phi_0  =  \exp (-2 \phi_{x})$,
$d_1$ and $r_x$. Other coefficients can be found from the following conditions:
\begin{eqnarray}\label{eq2.14}
\phi_1  =  \frac{m^2_{pl} [ d_1 (1 + r_x) - \sigma^2_0]}
{4 \phi_0 \lambda d_1}, \qquad
\phi_2   =  \frac{ - m^2_{pl} r_x \sigma_1 + 4 \phi_0 \lambda \phi_1 \sigma_1}
{6 \phi_0 \lambda \sigma_0}, \qquad
d_2  =  \frac{2}{3} \frac{d_1 \sigma_1}{\sigma_0}. \nonumber
\end{eqnarray}
``Curvature invariant'' $R_{ijkl}R^{ijkl} \sim
\frac{1}{x} + O (x^{\frac{1}{2}})) \rightarrow \infty$
and this means that $r_x$ represents the ``singular horizon''.
The distance between $r_x$ and $r_h$ is rather long for the
big values of $r_h$ and decreases with decreasing $r_h$.
In the limit point, defined by the equation (\ref{eq2.7}),
all points pour together $r_h=r_s=r_x$ and the solutions
inside $r_h$ do not exist. This can be seen from Fig.2.

\section{Thermodynamical properties of Gauss-Bonnet solution}

If one works in Loretzian spacetime, he can say nothing
about thermodynamical properties of a solution obtained.
Using the Euclidean version of the metric:
$$ds^2=\Delta d\tau^2 + \frac{\sigma^2}{\Delta}dr^2 + r^2 d\Omega,$$
where $\tau$ is a periodic coordinate which range is from $0$ to $\pi$
\cite{gibbons,ellis,liberati}, one can easily write the
inverse temperature \cite{kallosh}:
\begin{eqnarray}
\beta & = & 4 \pi \sqrt{g_{00} g_{11}}
\biggl[ \frac{d}{dr} g_{00} \biggr]^{-1} \mid_{r=r_h}.
\end{eqnarray}
>From equations (\ref{eq2.6})-(\ref{eq2.62})
it is possible to find that
$\beta=4 \pi (s_0 / d_1)$.

In the Euclidean frame the full action $S_E$ after integrating over periodic time $\tau$
takes the following form:
\begin{eqnarray}\label{eq4.1}
 S_E  & = & \frac{1}{4} \beta \int dr
 \biggl[ m^2_{pl}
\bigl(
-\frac{\Delta'' r^2}{\sigma}
-4 \frac{\Delta' r}{\sigma}
+\frac{\Delta' \sigma' r^2}{\sigma^2}
-2 \frac{\Delta}{\sigma}
+4 \frac{\Delta \sigma' r}{\sigma^2}
+2 \sigma
- 2 (\phi')^2 \frac{\Delta r^2}{\sigma} \bigr) \nonumber \\
& + & 4 \mbox{e}^{-2\phi} \lambda
\bigl[ \frac{\Delta \Delta'}{\sigma^3} - \frac{\Delta'}{\sigma} \bigr]' \biggr].
\end{eqnarray}
The boundary term obtained as it was done in \cite{gibbons} has
the following form:
\begin{eqnarray}\label{eq4.2}
K = \frac{1}{4} \beta
\biggl[ m^2_{pl}
\biggl( \frac{\Delta' r^2}{\sigma} +
2 \frac{\Delta r}{\sigma} \biggr) +
4 \mbox{e}^{-2\phi} \lambda
\biggl( \frac{\Delta \Delta'}{\sigma^3} - \frac{\Delta'}{\sigma} \biggr) \biggr].
\end{eqnarray}
For a non-extreme black-hole the boundaries of the
spacetime manifold are set by the extreme values of the radius coordinate.
They are $r=r_h$ and $r=\infty$. After removal of the conical singularity,
the spacetime has the only boundary at $r=\infty$, because the black hole horizon
is not a spacetime border \cite{liberati}. So, we have to take into account
only $r_h$  value of boundary term $K$. It takes the following form:
\begin{eqnarray}
K-K_0 = (K-K_0)_{r_h}  =  \frac{1}{4} \beta \biggl[ m^2_{pl}
\bigl( \frac{d_1 r^2_h}{s_0} - 2 r_h \bigr) - 4 \phi_0 \lambda \frac{d_1}{s_0} \biggr]
\end{eqnarray}
in terms of the asymptotic solution on the event horizon (\ref{eq2.7}).
$K_0$ is the boundary contribution of the flat spacetime. In this case the Euclidean
action $I_E$ takes the following form:
\begin{eqnarray}\label{eq4.5}
I_E = \int\limits_{r_h}^{\infty} dr S_E - (K-K_0)_{r_h},
\end{eqnarray}
where $S_E$ is defined by (\ref{eq4.2}).

The physical entropy can be found by the following expression \cite{mignemi}:
\begin{eqnarray}\label{eq4.7}
S(r_h) & = & \beta \frac{\partial I}{\partial \beta} - I_E
 =  \beta (r_h) \biggl(\frac{\partial I(r_h)}{\partial r_h} \biggr)
\biggl(\frac{\partial \beta(r_h)}{\partial r_h} \biggr)^{-1} - I_E(r_h).
\end{eqnarray}
Plots for the GB and Schwarzshild entropy are shown in Fig.3.

\section{Discussion and conclusions}

In this paper there are suggested
the black hole solutions with non-trivial dilaton
``hair'' of low energy effective SuperString Theory with second order
curvature corrections which are obtained independently from
Kanti et all \cite{kanti} and were found
outside and inside the event horizon by the
numerical method described in Appendix.
The solutions are characterized by ADM mass $M$, dilaton charge $D$ and
asymptotic dilaton value $\phi_\infty$. They are stable under the fluctuations
of initial conditions. So, we can arrive to a conclusion that
the SuperString Theory provides such situations by itself as can not be
covered by the ``no-hair`` theorem \cite{kanti}. As these solutions have
the non-perturbative nature, as they have no limits with
the perturbative parameter values.

The most interesting result of our work is the existence of $r_s$-singularity
inside the black hole after adding the GB term to the action.
This singularity has the topology $S^2 \times R^1$, i.e. it is an
infinite (in time direction) ``tube'' of radius $r_s$. Similar ``tube''
in the Schwarzshild metric with an additional condition of
$R^{abcd}R_{abcd}$ finiteness was discussed by V.Frolov et al.
\cite{frolov}.
There are two solutions on this ``tube''. The asymptotically flat solution, which
is the main one, starts from $r_s$ and continues without limits up to the infinity.
In the case of $r_h$ to be quite large and $r_h \gg r_s$ it is possible
to suppose $r_s$ to be approximately vanished, then the main solution looks
like the standard Schwarzshild one with a constant dilatonic field (see Fig.2)
which agrees with results of Mignemi \cite{mignemi} and Kanti \cite{kanti}.
The additional solution branch provides the existence of the ``singular''
inner horizon with $R_{ijkl}R^{ijkl} \rightarrow \infty$.
Some solutions exist inside ``tube'' $r_s$, but they
are unstable under the initial condition fluctuations, and we
can not distinguish which branch --- the main one or the additional one ---
they correspond to.
One can suppose  that this $r_s$ singularity is the
consequence of the bad metric parametrization choice.
In order to prove (or not)
this artifact, the solution in  another metric was studied. This is the metric of such class
that was suggested by P.Breitenlohner et al. \cite{breit} and
has the following form:
\begin{eqnarray}
ds^2 = \beta^2 (\tau) dt^2 - d \tau^2 - r^2 (\tau) d \Omega^2 .
\end{eqnarray}
Topology $S^2 \times R^1$ inside the black hole horizon also exists
in this metric. So the presence of $r_s$ ``tube'' is not the consequence
of the metric parametrization choice.
The question on the more detail structure of the ``tube'' $r_s$
is the subject of the next study.

The Gauss-Bonnet term influence to the solution behavior
was also considered. When the horizon radius $r_h$ is quite
large or $\lambda$ becomes quite small the main contribution
to the action comes from the Einstein part ($r_h \gg r_s$ $\Rightarrow$
$r_s \simeq 0$ in comparison with $r_h$). While decreasing $r_h$
(increasing $\lambda$), the GB influence increases,
the dilaton charge absolute value becomes larger, the distance between $r_s$
and $r_h$ becomes smaller. Once the limit position is achieved where
$r_h=r_s=r_x \equiv r_{hxs}$. This is a minimal
point which provides the solution
like the black hole type. In the point $r_{hxs}$ the solution (main branch)
exists only outside the event horizon.

The Euclidean action and the entropy behavior of Einstein--dilaton--GB
solution was also examined. It was found that the inverse temperature
is not far from the corresponding Schwarzshild one. When
$r_h$ becomes quite large, the entropy $S$ is quite equal to the
Schwarzshild one $S_{SW}$. During decreasing $r_h$, the difference between $S$
and $S_{SW}$ becomes larger. $S$ has always the positive signature.
This conclusion does not contradict to Mignemi's formula \cite{mignemi},
obtained by the perturbative method in the case of rather large $r_h$,
which has the following form in our variables:
\begin{equation}\label{eq4.9}
S = 4 \pi M^2 \biggl( 1 + \lambda \frac{1}{M^2} +
\lambda^2 \frac{73}{120}\frac{1}{M^4}\biggr).
\end{equation}
Our entropy value (\ref{eq4.7}) was obtained with the non-perturbative method.
So, it is possible to note, that the formula (\ref{eq4.9})
is indeed correct with the quite large values of $r_h$ and gives
the increasing mistake if
entropy becomes small.

\acknowledgments
One of ther authors (S.A.) would like to thank Professor D.V.Gal'tsov
for useful discussions on the subject of this work.

\section{Appendix: The method of integration over a parameter of ordinary
differential equations given in a non-evident form}

Let us consider a system of usual differential equations
with unknown functions $x_i (t) \in R^n$ which are not solved
relatively derivatives $x'_i (t)$ to have the following form:
\begin{eqnarray}{\label{a.1}}
&& \sum\limits_{j=1}^n x'_j(t) a_{1j}(x,t) = b_1(x,t), \nonumber \\
&& \sum\limits_{j=1}^n x'_j(t) a_{2j}(x,t) = b_2(x,t), \nonumber \\
&& \qquad\qquad\cdots \\
&& \sum\limits_{j=1}^n x'_j(t) a_{nj}(x,t) = b_n(x,t), \nonumber
\end{eqnarray}
where the coefficients $b_i$, $a_{ij} \in C^1 [R^n \times R^1]$.
For the simplicity it is convenient to introduce $(n \times n)$
matrix \\
$A(x,t) \equiv \{ a_{ij}(x,t)\}_{i,j=1,\ldots, n}$.
The system (\ref{a.1}) is a linear $x'_i (t)$ system, therefore, it can be solved
and its solution has the following matrix form:
\begin{eqnarray}{\label{a.2}}
x'(t)=A^{-1} (x,t) b^{\dagger}(x,t),
\end{eqnarray}
where transposed vector $b^{\dagger}(x,t)$ is equal to
$(b_1(x,t),b_2(x,t),\ldots, b_n(x,t))^{\dagger}$.
Let $t_0$ to be the unit value of $t$ and $x(t_0)=x_0$. So, one has the
Cauchy problem that is to continue the solution of
(\ref{a.1}) to the manifold $t \in [t_0, t_1 ]$.
According to the existence theorem, the solution $x(t, t_0, x_0)$
of (\ref{a.1}) would exist only in the case of the main system discriminant
to be not equal to zero on the solutions of (\ref{a.1}),
beginning from the point $(x_0, t_0)$, i.e.:
\begin{eqnarray}{\label{a.3}}
\det A (x(t,t_0,x_0),t) \neq 0.
\end{eqnarray}
If in any point $t^*$ the following condition is realized, i.e.
$\det A (x(t,t_0,x_0),t) = 0$, the system (\ref{a.1})
has a non-single solution, and a numerical integration performed
by the usual Runge-Kutta, Adams, ets. method will stop before the point
$t^*$. So, it is necessary to choose such a numerical integration strategy
which allows to pass the point $t^*$. The same method
used for other problem classes was discussed early, for example,
by V.A.Egorov et all \cite{egorov}.

For the purposes discussed in the paper one can introduce  a new independent
variable $s$ and further to consider $x$ and $t$ to be functions of $s$:
$x=x(s)$, $t=t(s)$. Therefore,
\begin{eqnarray}{\label{a.4}}
\frac{dx}{dt} = \frac{dx}{ds} \frac{ds}{dt} = x'_s (t'_s)^{-1}.
\end{eqnarray}
Putting (\ref{a.4}) to the system (\ref{a.1}), one obtains
a new homogeneous system which consists of $n$ equations with $(n+1)$
unknowns. One needs to add $(n+1)$th equation representing
the normalization condition to the system, which takes
the following form:
\begin{eqnarray}{\label{a.5}}
&& \sum\limits_{j=1}^n \frac{dx_j}{ds} a_{1j}(x,t) -
\frac{dt}{ds} b_1(x,t) = 0, \nonumber \\
&& \sum\limits_{j=1}^n \frac{dx_j}{ds} a_{2j}(x,t) -
\frac{dt}{ds} b_2(x,t) = 0, \nonumber \\
&& \qquad\qquad\cdots \\
&& \sum\limits_{j=1}^n \frac{dx_j}{ds} a_{nj}(x,t) -
\frac{dt}{ds} b_n(x,t) = 0, \nonumber \\
&& \sum\limits_{j=1}^n \frac{dx_j}{ds} \tau_j (s) -
\frac{dt}{ds}\tau_t (s) = 1, \nonumber
\end{eqnarray}
where new functions $\tau (s) \in R^{n+1}$. These new functions $\tau (s)$
are fixed by the following normalization conditions:
\begin{eqnarray}{\label{a.6}}
 \tau_i (s)  =  \frac{dx_i/ds}{\sqrt{\sum\limits_{j=1}^n (dx_i/ds)^2 +
(dt/ds)^2}}, \quad
 \tau_t (s)  =  \frac{dt/ds}{\sqrt{\sum\limits_{j=1}^n (dx_i/ds)^2 +
(dt/ds)^2}},
\end{eqnarray}
So, $\vec{\tau}(s)$ is the tangent vector to the solution trajectory in $(n+1)$
phase-space, therefore, the integration is proceeded along the
solution trajectory, not along the
radial coordinate $r$, as usual performed.
At a first step the most convenient choice of $\vec{\tau}$ is the following:
$\tau_i=0$, $\tau_t=\pm 1$, where its sign depends upon
the solution direction moving.
At the next step $\vec{\tau}(s)$ are extrapolated by the Legendre polynoms
using the $\vec{\tau}(s)$ values from the previous steps according to
formulae (\ref{a.6}). Therefore the system (\ref{a.5}) remains the linear one.
One can argue that if the integrating step vanishes the solution
of the system (\ref{a.5})
with the extrapolating functions $\vec{\tau}(s)$ reduce
to the solution of the system (\ref{a.5}) with the functions
$\vec{\tau}(s)$ defined in (\ref{a.6}).
In such case our system
(\ref{eq2.1}), (\ref{eq2.3}), (\ref{eq2.4})
is written as follows:
\begin{eqnarray}{\label{a.7}}
\begin{array}{ccccccccccccc}
\frac{d\Delta}{ds} & + & 0 & + & 0 & + & 0 & + & 0 &
- & \Delta' \frac{dr}{ds} & = & 0, \\
\\
0 & + & \frac{d\phi}{ds} & + & 0 & + & 0 & + & 0 & -
& \phi' \frac{dr}{ds} & = & 0, \\
\\
0 & + & 0 & + & a_{11} \frac{d\Delta'}{ds} & +
& a_{12} \frac{d\sigma}{ds} & + & a_{13} \frac{d\phi'}{ds} & -
& b_1 \frac{dr}{ds} & = & 0, \\
\\
0 & + & 0 & + & a_{21} \frac{d\Delta'}{ds} & +
& a_{22} \frac{d\sigma}{ds} & + & a_{23} \frac{d\phi'}{ds} & -
& b_2 \frac{dr}{ds} & = & 0, \\
\\
0 & + & 0 & + & a_{31} \frac{d\Delta'}{ds} & +
& a_{32} \frac{d\sigma}{ds} & + & a_{33} \frac{d\phi'}{ds} & -
& b_3 \frac{dr}{ds} & = & 0, \\
\\
\frac{d\Delta}{ds} \tau_\Delta & +
& \frac{d\phi}{ds} \tau_\phi & + & \frac{d\Delta'}{ds} \tau_{\Delta'} & +
& \frac{d\sigma}{ds} \tau_{\sigma} & +
& \frac{d\phi'}{ds} \tau_{\phi'} & - & \tau_r \frac{dr}{ds} & = & 1,
\end{array}
\end{eqnarray}
where stroke denotes $\partial / \partial r$ and $a_{ij}$, $b_i$ have the
form:
\begin{eqnarray}
a_{11} & = & 0, \nonumber \\
a_{12} & = & - m_{pl}^2 \sigma^2 r + 4 \mbox{e}^{-2 \phi} \lambda \phi'
(\sigma^2 - 3 \Delta ), \nonumber \\
a_{13} & = & 4 \mbox{e}^{-2 \phi} \lambda \sigma (\Delta - \sigma^2), \nonumber \\
a_{21} & = & m_{pl}^2 \sigma^3 r + 4 \mbox{e}^{-2 \phi} \lambda \phi' 2 \Delta \sigma , \nonumber \\
a_{22} & = & - m_{pl}^2 \sigma^2 (\Delta' r + 2 \Delta)
- 4 \mbox{e}^{-2 \phi} \lambda \phi' 2 \Delta 3 \Delta' ,\nonumber \\
a_{23} & = & 4 \mbox{e}^{-2 \phi} \lambda 2 \Delta  \Delta' \sigma ,\nonumber \\
a_{31} & = & 4 \mbox{e}^{-2 \phi} \lambda (\Delta\sigma - \sigma^3) ,\nonumber \\
a_{32} & = & 2 m_{pl}^2 \sigma^2 \Delta r^2 \phi'
+ 4 \mbox{e}^{-2 \phi} \lambda (-3 \Delta \Delta' + \Delta' \sigma^2) ,\nonumber \\
a_{33} & = & - 2 m_{pl}^2 \sigma^2 \Delta r^2 \sigma ,\nonumber \\
\nonumber \\
b_{1} & = & - m_{pl}^2 \sigma^3 r^2 (\phi')^2 +
4 \mbox{e}^{-2 \phi} \lambda \sigma (\Delta  - \sigma^2) 2 (\phi')^2 ,\nonumber \\
b_{2} & = & - m_{pl}^2 \sigma^2 (2 \Delta' \sigma + 2 \Delta r \sigma (\phi')^2 ) +
4 \mbox{e}^{-2 \phi} \lambda 2 \sigma \Delta \Delta' 2 (\phi')^2 -
4 \mbox{e}^{-2 \phi} \lambda \phi' 2 (\Delta')^2 \sigma ,\nonumber \\
b_{3} & = & 2 m_{pl}^2 \sigma^2 (\Delta' r^2 \sigma \phi' + 2 \Delta r \sigma \phi' )
- 4 \mbox{e}^{-2 \phi} \lambda   (\Delta')^2 \sigma .\nonumber
\end{eqnarray}

The integration is proceeded by the independent variable quantity $s$
and starts from the point of $s=0$. At the first step it is necessary
the following condition to be fulfilled: $\det B $=$\det A \neq 0$, where $B$
is the equation matrix defined as follows:
\begin{eqnarray}{\label{a.8}}
B=\left(
\begin{array}{cccccc}
1 & 0 & 0 & 0 & 0 & - \Delta' \\
0 & 1 & 0 & 0 & 0 & - \phi' \\
0 & 0 & a_{11} & a_{12} & a_{13} & - b_1 \\
0 & 0 & a_{21} & a_{22} & a_{23} & - b_2 \\
0 & 0 & a_{31} & a_{32} & a_{33} & - b_3 \\
\tau_{\Delta} & \tau_{\phi} & \tau_{\Delta'}
& \tau_{\sigma} & \tau_{\phi'} & \tau_{r}
\end{array}
\right)
\end{eqnarray}
At every integration step one calculates $\vec{\tau}$
by the formulae (\ref{a.6}). The 8th order Runge-Kutta program
for integration by one step was
used in our work. This program divides the given step automatically
to reach the accuracy required.

One can argue that if the integration step
becomes quite small, the matrix $B$ (\ref{a.8}) will
not be degenerated at that trajectory point
and rang$ A \geq (n-1)$. When rang$ A < (n-1)$, a ramifying point
for two or more branches can exist in that point.
Such a point can not be passed by the describing
strategy.

Matrix $B$ can be turned out
by the Singular Value Decomposition (SVD) method \cite{nag}.
It decomposes the given matrix $B$ to the product of three matrixes
$U \Sigma V^{\dagger}$, where $U$ and $V$ are the unitary matrixes
$(n+1)\times(n+1)$ dimension and $\Sigma$ is the diagonal matrix of absolute own
values of $B$, i.e. $\Sigma = \{ \sigma_{ii} \}$, $\sigma_{ii} \geq 0$.
Therefore, $B^{-1}=V \Sigma^{-1} U^{\dagger}$ where $\Sigma^{-1}=\{1/ \sigma_{ii} \}$.
The most effective SVD algorithm can be found in the Numerical Analytical Group (NAG).
It is very useful because during the calculation one sees
the degeneracy's order of matrix $B$ (singular values $\sigma_{ii}$ vanish).

In the calculations it is desirable to work with Plank unit values,
setting $m_{pl}=1$. For simplicity the parameter $\lambda$ in the action (\ref{eq1})
was also set to be equal to one. Equations (\ref{eq2.9}),
(\ref{eq4.5}) were added to the main system (\ref{a.7})
to reach the maximum available accuracy for calculating ADM mass $M$,
dilaton charge $D$ and Euclidean action $I_E$.

\newpage

\begin{figure}
\caption{The dependence of the metric functions $\Delta$ (a), $\sigma$ (b)
and dilaton function $\exp(-2\phi)$ (c) versus the radial coordinate $r$
when the event horizon value $r_h$ is equal to $10.0$.}
\end{figure}
\begin{figure}
\caption{3D plots of
the dependence of the metric function $\Delta (r)$ (a)
and the dilaton function $\exp(-2\phi(r))$ (b) versus
the event horizon value $r_h$.}
\end{figure}
\begin{figure}
\caption{The entropy behavior of the Schwarzshild solution (line)
and GB one (squares)
versus
the event horizon value $r_h$.}
\end{figure}

\newpage

\begin{table}
\caption{The dependence of ADM mass $M$, dilaton charge $D$,
GB inverse temperature $\beta$ and Schwarzshild one $\beta_{SW}$
versus the event horizon value $r_h$.}
\begin{tabular}{|c|c|c|c|c|c|c|c|}\hline
$r_h$   &   3.5   &   5.0   &   10.0   &   15.0   &  20.0    & 30.0     & 100.0    \\ \hline
$M$     & 1.8531  & 2.5365  & 5.0048   & 7.5014   & 10.0006  & 15.0001  & 50.0000  \\ \hline
$D$     & -0.4658 & -0.3632 & -0.1951  & -0.1319  & -0.0994  & -0.0665  & -0.01999 \\ \hline
$\beta$ & 43.1675 & 62.8607 & 125.6630 & 188.4898 & 251.2456 & 376.9911 & 1256.6370 \\ \hline
$\beta_{SW}$ & 43.9824 & 62.8320 & 125.6640 & 188.4960 & 251.3280 & 376.9920 & 1256.6400 \\ \hline
\end{tabular}
\end{table}

\end{document}